\documentclass[prd,aps,twocolumn,preprintnumbers, showpacs, nofootinbib,superscriptaddress,notitlepage]{revtex4-1}
\usepackage{amssymb, amsthm}
\usepackage{amsmath}    
\usepackage{mathrsfs}   
\usepackage{graphicx}   
\usepackage{xcolor}     
\usepackage{footnote}   
\usepackage{slashed}    
\usepackage{subfigure}  
\usepackage{hyperref}   
\usepackage{rotating}   
\usepackage{multirow}   
\usepackage[normalem]{ulem} 
\usepackage[utf8]{inputenc}

\newcommand{\beq}{\begin{eqnarray}}
\newcommand{\eeq}{\end{eqnarray}}

\begin{document}

\preprint{MSUHEP-18-013,MIT-CTP/5032}

\title{Proton Isovector Helicity Distribution on the Lattice at Physical Pion Mass}

\collaboration{\bf{$\rm {\bf LP^3}$ Collaboration}}

\author{Huey-Wen Lin}
\affiliation{Department of Physics and Astronomy, Michigan State University, East Lansing, MI 48824, USA}
\affiliation{Department of Computational Mathematics, Michigan State University, East Lansing, MI 48824, USA}

\author{Jiunn-Wei Chen}
\affiliation{Department of Physics, Center for Theoretical Physics, and Leung Center for Cosmology and Particle Astrophysics, National Taiwan University, Taipei 106, Taiwan}
\affiliation{Center for Theoretical Physics, Massachusetts Institute of Technology, Cambridge, MA 02139, USA}

\author{Xiangdong Ji}
\affiliation{Tsung-Dao Lee Institute and School of Physics and Astronomy,
Shanghai Jiao Tong University, Shanghai, 200240, China}
\affiliation{Maryland Center for Fundamental Physics,
Department of Physics, University of Maryland,
College Park, Maryland 20742, USA}

\author{Luchang Jin}
\affiliation{Physics Department, University of Connecticut,
Storrs, Connecticut 06269-3046, USA}
\affiliation{RIKEN BNL Research Center, Brookhaven National Laboratory,
Upton, NY 11973, USA}

\author{Ruizi Li}
\affiliation{Department of Physics and Astronomy, Michigan State University, East Lansing, MI 48824, USA}

\author{Yu-Sheng Liu}
\email{Corresponding author: mestelqure@gmail.com}
\affiliation{Tsung-Dao Lee Institute, Shanghai Jiao-Tong University, Shanghai 200240, China}

\author{Yi-Bo Yang}
\email{Corresponding author: ybyang@itp.ac.cn}
\affiliation{Department of Physics and Astronomy, Michigan State University, East Lansing, MI 48824, USA}
\affiliation{Institute of Theoretical Physics, Chinese Academy of Sciences, Beijing 100190, China}

\author{Jian-Hui Zhang}
\affiliation{Institut f\"ur Theoretische Physik, Universit\"at Regensburg, D-93040 Regensburg, Germany}

\author{Yong Zhao}
\affiliation{Center for Theoretical Physics, Massachusetts Institute of Technology, Cambridge, MA 02139, USA}


\begin{abstract}
We present a state-of-the-art calculation of the isovector quark helicity Bjorken-$x$ distribution
in the proton using lattice-QCD ensembles at the physical pion mass. We compute quasi-distributions at proton momenta
$P_z \in \{2.2, 2.6, 3.0\}$~GeV on the lattice, and match them systematically to the physical parton distribution using
large-momentum effective theory (LaMET). We reach an unprecedented precision through high statistics in simulations,
large-momentum proton matrix elements, and control of excited-state contamination.
The resulting distribution with combined statistical and systematic errors is in agreement with the latest phenomenological analysis of the
spin-dependent experimental data; in particular, $\Delta \bar{u}(x)>\Delta \bar{d}(x)$.
\end{abstract}

\maketitle

Understanding the spin structure of the proton is a challenging frontier problem in modern physics.
Some of the most studied physical observables are the parton helicity distributions $\Delta q(x)$ and $\Delta g(x)$,
which describe the number densities of polarized partons (quarks and gluons) with momentum fraction $x$
in a longitudinally polarized proton.
Decades of polarized deep inelastic scattering (DIS) and semi-inclusive DIS (SIDIS)
data at a wide range of kinematics have greatly improved our knowledge of these distributions.
Significant progress has also been made in recent years in polarized proton-proton collisions at the Relativistic Heavy-Ion Collider (RHIC).
Groups such as DSSV14~\cite{deFlorian:2014yva}, NNPDFpol1.1~\cite{Nocera:2014gqa}, and JAM17~\cite{Ethier:2017zbq}
have used the available experimental data to yield the phenomenological helicity-dependent distributions.
In the future, the kinematic coverage for spin-dependent parton distribution functions (PDFs) is expected to be greatly expanded with new
data on DIS and SIDIS from Jefferson Lab 12-GeV~\cite{Dudek:2012vr} and a future Electron-Ion Collider (EIC)~\cite{Accardi:2012qut}.

Lattice gauge theory allows {\it ab initio} calculations of the proton spin structure from the fundamental theory of strong interaction:
quantum chromodynamics (QCD). The lowest moments of the polarized quark distribution are matrix elements of local operators, and have
been studied extensively using lattice calculations (see~\cite{Lin:2017snn} for a review). On the other hand,
$x$-dependent PDFs have until recently defied theoretical attempts, fundamentally
because PDFs are defined through the matrix elements of lightcone correlations, whereas the lattice approach
is intrinsically Euclidean. Large-momentum effective theory~(LaMET)~\cite{Ji:2013fga,Ji:2013dva,Hatta:2013gta,Ji:2014gla,Ji:2014lra} recently provided a breakthrough in calculation of the $x$-dependence of PDFs using lattice QCD.
On the lattice, one can calculate the matrix elements of Euclidean observables in a large-momentum hadron
state (often called ``quasi-PDFs'' in the study of parton distributions), which can be used to extract the nonperturbative lightcone dynamics
through factorization and matching.

There has been much progress in the last few years in applying LaMET to calculate lightcone physics (see Ref.~\cite{Liu:2018uuj}) for
a more complete list of references). In particular, the renormalization properties of the quasi-PDF
operators and nonperturbative renormalization (NPR) on lattice have been understood and implemented ~\cite{Ji:2015jwa,Ishikawa:2016znu,Chen:2016fxx,Constantinou:2017sej,Ji:2017oey,Ishikawa:2017faj,Green:2017xeu,Chen:2017mzz,Alexandrou:2017huk}.
Progress also has been made in studying spin-dependent lightcone physics in LaMET. Our pioneering exploratory calculation on quark helicity PDFs~\cite{Lin:2014zya} was done at pion mass $M_\pi \approx 310$~MeV, with the largest proton momentum around 1.3~GeV.
A later calculation by ETMC~\cite{Alexandrou:2016jqi} at slightly heavier pion mass showed similar results.
The full matching calculations and mass corrections were reported in~\cite{Chen:2016utp}.
Helicity-distribution calculations in regularization-independent momentum-subtraction (RI/MOM) scheme and NPR at physical pion mass were first reported by us~{\cite{Lin:2017ani}}, and more recently with high
statistics by ETMC~\cite{Alexandrou:2018pbm}.

In this paper, we report a state-of-the-art calculation at physical pion mass on the isovector quark helicity PDF, $\Delta u(x)-\Delta d(x)$,
in the proton. Large-momentum (up to 3.0~GeV) proton sources have been employed to suppress high-twist contributions to quasi-PDFs.
The proton matrix elements are renormalized in RI/MOM scheme, along
with a matching formula to connect the RI/MOM quasi-PDF to the physical PDF in $\overline{\text{MS}}$ scheme~\cite{Stewart:2017tvs,Liu:2018uuj}.
Six source-sink separations in combination with multiple-state analysis help to remove excited-state contamination from the proton state.
In the moderate- to large-$x$ region, the final result with combined statistical and systematic errors shows a significant improvement compared to previous lattice studies and is consistent with the global analyses by NNPDF and JAM groups.
We also see evidence that $\Delta \bar{u}(x)>\Delta \bar{d}(x)$, as found in experimental data.



To calculate the quark helicity PDFs in LaMET, we start by computing a quasi-PDF on a lattice with spacing $a$,
\begin{equation}\label{eq:quasipdf}
\Delta\tilde{q}(x, P_z, a)
  = \int_{-\infty}^\infty \frac{P_z dz}{2\pi}\ e^{ixP_zz} \frac{1}{2 P_0} \big\langle PS \big| \hat O(z,a) \big|PS\big\rangle
\end{equation}
where $P_\mu=(P_0,0,0,P_z)$ and $S_\mu=(P_z,0,0,P_0)$ are the proton four-momentum and
longitudinal polarization vectors, respectively. The nonlocal Euclidean operator is
$\hat O(z,a)=\bar{\psi}_q(z)\gamma^z\gamma_5 U(z, 0) \psi_q (0)$ with the Wilson line $U(z, 0)= P\exp\left(-ig\int_0^z dz' A_z(z')\right)$ and
subscript $q=(u,d,s,...)$ as a flavor index. Here, we consider the isovector combination, $\Delta \tilde u-\Delta \tilde d$,
so that the disconnected contributions on lattice cancel.

$\hat O(z,a)$ has both power and logarithmic divergences as $a\to 0$, and for the isovector
combination, all divergences have been shown to factorize~\cite{Ji:2017oey,Ishikawa:2017faj,Green:2017xeu}.
To achieve high precision in matching, a NPR for the lattice operators is used to define the continuum limit of the
quasi-PDF matrix elements. Following the RI/MOM scheme advocated in Refs.~\cite{Stewart:2017tvs,Chen:2017mzz},
we introduce a $z$-dependent renormalization factor ${Z}(z,p^R_z,\mu_R,a)$ defined on the lattice
in an off-shell quark state in Landau gauge with $z$-component momentum $p_z^R$ and subtraction
scale $\mu_R$.
The renormalized matrix element of $\tilde{h}(z,P_z,a) = ({1}/{2P_0}) \langle PS |\hat O(z,a)| PS
\rangle$ in coordinate space,
\begin{align} \label{eq:rimomh}
\tilde{h}_R(z,P_z, p^R_z,\mu_R)=&{Z}^{-1}(z,p^R_z,\mu_R,a)\tilde{h}(z,P_z,a) ,
\end{align}
has a well defined continuum limit as $a\to 0$.

Following the framework described in
 Refs.~\cite{Stewart:2017tvs,Izubuchi:2018srq}, the matching between the renormalized
 quasi-PDF $\Delta \tilde{q}_R(x,P_z, p^R_z,\mu_R)$ and the physical PDF $\Delta
 q(y,\mu)$ at scale $\mu$ is
\begin{align}
\label{eq:momfact}
\Delta \tilde{q}_R(x,P_z, p^R_z,\mu_R) =& \int_{-1}^1 {dy\over |y|}\: C\left({x\over y},r,\frac{yP_z}{\mu},\frac{yP_z}{p_z^R}\right) \, \Delta q(y,\mu)\nonumber\\
+&\mathcal{O}\left({M^2\over P_z^2},{\Lambda_{\text{QCD}}^2\over P_z^2}\right),
\end{align}
where $r={\mu_R}^2/{(p^R_z)}^2$, $M$ is the proton mass,
and the antiquark distribution $\Delta \overline{ q}(y,\mu)\equiv \Delta q(-y,\mu)$ falls in the region $-1<y<0$. The matching coefficient $C$ at one-loop level using minimal projection in the $\overline{\text{MS}}$ scheme can be found in Ref.~\cite{Liu:2018uuj}.

We perform lattice calculations of the bare isovector quark helicity quasi-PDF using clover valence fermions on an ensemble of $884$ gauge configurations with lattice spacing $a=0.09$~fm, box
size $L\approx 5.8$~fm, and with pion mass $M_\pi \approx 135$~MeV and $N_f=2+1+1$ (degenerate up/down, strange and charm) flavors of highly improved staggered dynamical quarks (HISQ)~\cite{Follana:2006rc} generated by MILC Collaboration~\cite{Bazavov:2012xda}. The gauge links are one-step hypercubic (HYP)-smeared~\cite{Hasenfratz:2001hp} to suppress the discretization effects. The clover parameters are tuned to recover the lowest pion mass of the staggered quarks~\cite{Rajan:2017lxk,Bhattacharya:2015wna,Bhattacharya:2015esa,Bhattacharya:2013ehc}. We use multigrid algorithm~\cite{Babich:2010qb,Osborn:2010mb} in Chroma software package~\cite{Edwards:2004sx} to speed up the clover fermion inversion of
the quark propagator at physical pion mass, allowing a high-statistics calculation.

We use Gaussian momentum smearing~\cite{Bali:2016lva} for the quark field
$\psi(x)
+ \alpha \sum_j U_j(x)e^{ik\hat{e}_j}\psi(x+\hat{e}_j)$,
where $k=6$ 
is the input momentum parameter,
$U_j(x)$ are the gauge links in the $j$ direction, and
$\alpha$ is a tunable parameter as in traditional Gaussian smearing.
Such a momentum smearing is designed to increase overlap of the lattice sources with the ground-state proton of the desired momenta,
which allows us to reach higher-momentum states than was previously possible~\cite{Lin:2017ani}. This calculation employs
sources with $\vec{P}=\{0,0, n \frac{2\pi}{L}\}$, with $n \in \{10,12,14\}$, which correspond to 2.2, 2.6 and 3.0~GeV proton momenta, respectively.

\begin{figure}[htbp]
\includegraphics[width=.48\textwidth]{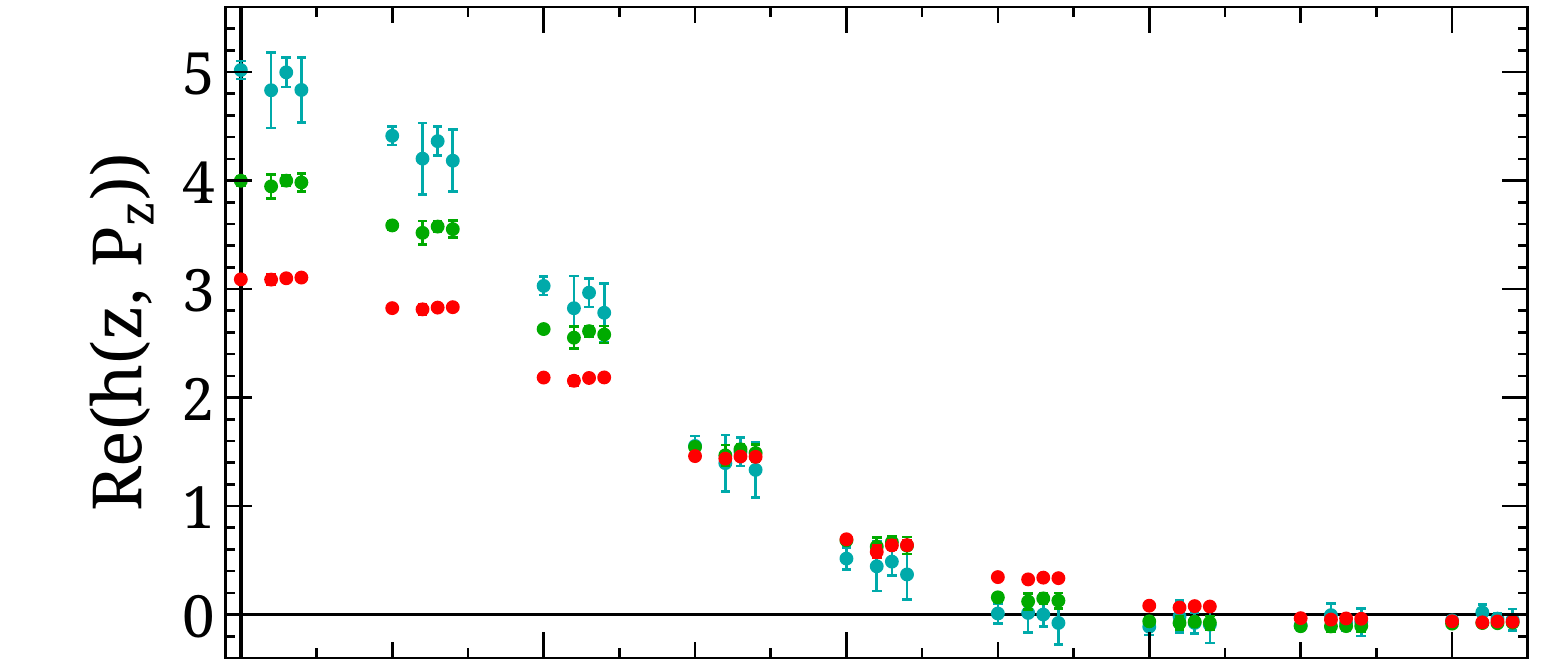}
\includegraphics[width=.48\textwidth]{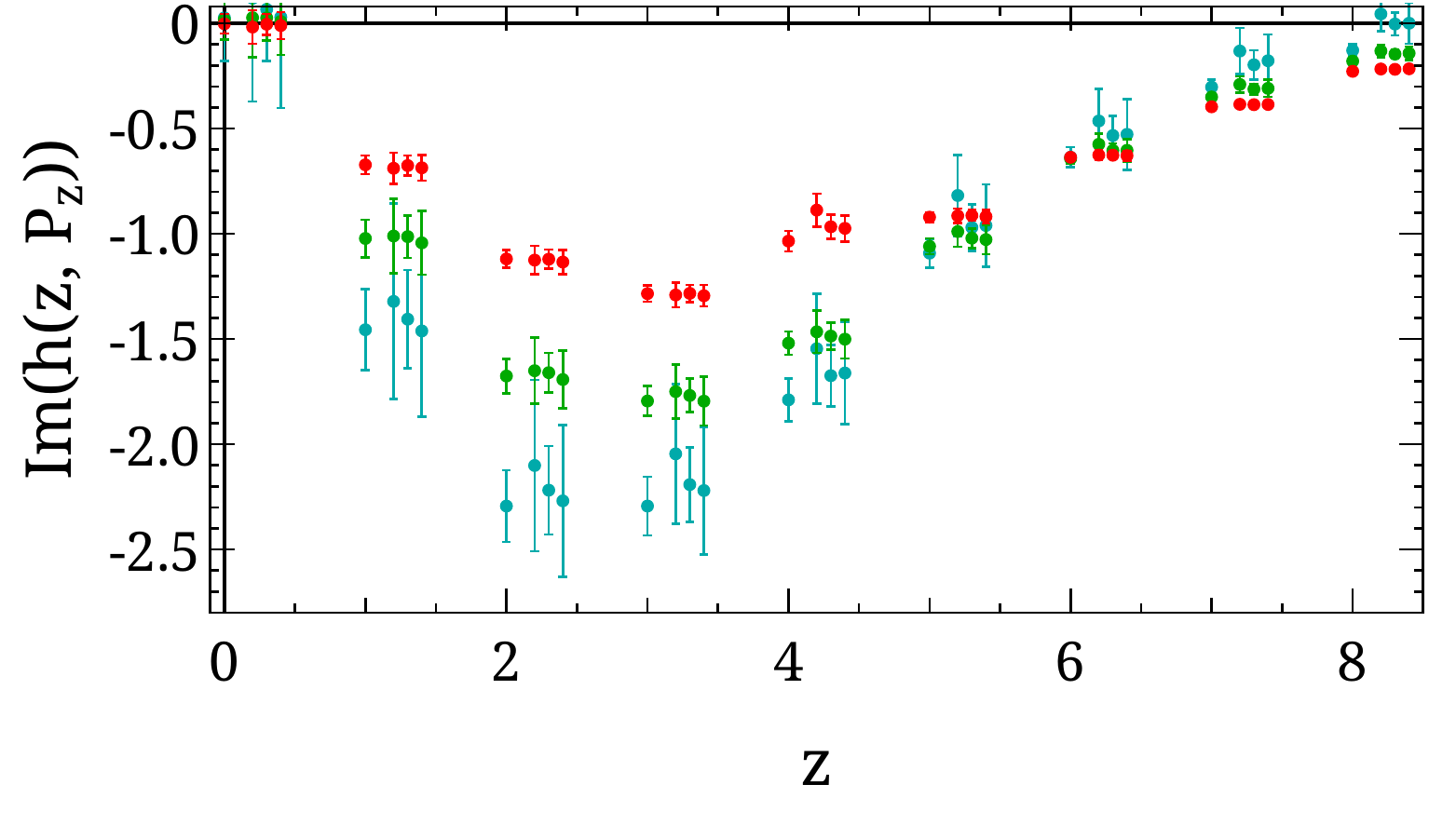}
\caption{
The real (top) and imaginary (bottom) parts of the bare proton matrix elements for the isovector quark helicity as functions of $z$ at all three momenta (2.2, and 2.6 and 3.0~GeV indicated by red, green and blue, respectively). Their kinematic factors have been omitted to enhance visibility by separating the small-$z$ matrix elements. At a given positive $z$ value, the data are slightly offset to show different ground-state extraction strategies; from left to right they are: two-simRR using all $t_\text{sep}$ (Fit-1), two-simRR using the largest 5 $t_\text{sep}$ (Fit-2), two-sim using the largest 4 $t_\text{sep}$ (Fit-3), and two-sim using the largest 3 $t_\text{sep}$ (Fit-4). All fits yield consistent results, as would be expected if the excited-state contamination is well-described by the two-state model. }
\label{fig:bareME-tsep}
\end{figure}

We investigate the excited-state contamination in the proton matrix elements by fitting data with different source-sink separations.
As the proton momentum increases, we anticipate stronger excited-state contamination since the excitation spectrum gets
compressed.  We measure the proton matrix elements with six source-sink separations, $t_\text{sep} \in \{0.54,0.72,0.81,0.90,0.99,1.08\}$~fm with
the number of measurements $\{16,32,32,64,64,128\}$k, respectively.
We use four two-state fits~\cite{Bhattacharya:2013ehc} to remove excited-state systematics among these source-sink separations by varying the number of excited-state matrix elements (``two-sim'' and ``two-simRR'') and the smallest $t_\text{sep}$ in the analysis.
Fit-1 uses the ``two-simRR''analysis~\cite{Bhattacharya:2013ehc}, which includes two additional matrix elements related to excited states. To counter the increase of degrees of freedom, we use all six separations; the fit uses only the largest five separations as Fit-2.
Fit-3 uses the ``two-sim'' analysis (with only one additional excited-state related element) to obtain the ground-state nucleon matrix elements using largest four source-sink separations. Fit-4 uses the same strategy as in Fit-3 but with only the largest three source-sink separations.
Fig.~\ref{fig:bareME-tsep} shows the bare matrix elements for a range of positive $z$ for all three momenta;
all four fits yield consistent results. The two-simRR analysis using $t_\text{sep}$ as small as 0.54~fm (Fit-1) gives consistent results
with the two-sim analysis using $t_\text{sep,min}$ of 0.81~fm (Fit-3),  with approximately the same statistical errors after removing
the excited-state contamination. Similar results are obtained by two other fits Fit-2 and Fit-4,
except with larger uncertainty due to fewer three-point proton correlators.
We use the fit with two-simRR with $t_\text{sep,min}=0.72$~fm for  our final analysis.

\begin{figure}[htbp]
\includegraphics[width=.48\textwidth]{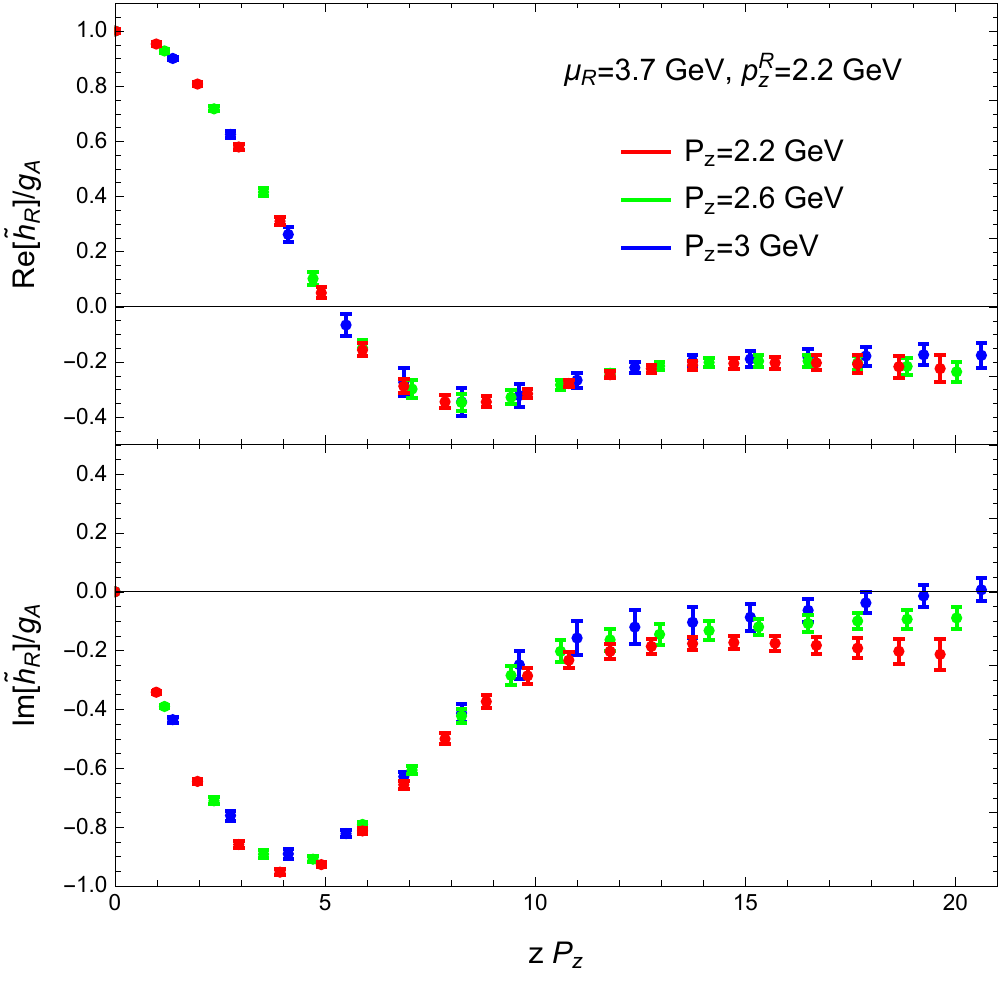}
\caption{The real (top) and imaginary (bottom) parts of the renormalized proton matrix elements as functions of $zP_z$, at renormalization scale
$\mu_R=3.7$ GeV, and $p_z^R =2.2$ GeV.
} \label{fig:ZME}
\end{figure}


To obtain the nonperturbative renormalization factor, one needs to calculate
the matrix elements of $\hat O(z,a)$ in a large-momentum quark state with point sources.
The momentum dependence is studied with the $z$-component ranging from $p^R_z=0$ to $3$~GeV at
off-shell mass $\mu_R=2.3$ and 3.7~GeV.
For $\mu_R=3.7$~GeV, the renormalization factor changes appreciably in the small-$p^R_z$ region, whereas at large $p^R_z$, it reaches a plateau.
Similar behavior is observed in the $\mu_R=2.3$~GeV case. We pick $p^R_z=2.2$~GeV as our
central value for the renormalization factor.

The renormalized isovector quark-helicity correlators as functions of $zP_z$ are shown in Fig.~\ref{fig:ZME}
for $\mu_R=3.7$~GeV, and $p_z^R =2.2$~GeV, with the real part shown in the top panel
and the imaginary at the bottom. The red, green, and blue colors indicate proton momenta of 2.2, 2.6 and 3.0~GeV,
respectively. We normalize all the matrix elements with $\tilde{h}_R (P_z,z=0)$ and multiply the final result
by $g_A=1.275$.  The nonzero long-range correlation in $zP_z$ reflects the significant
presence of small-momentum partons. The data indicate that the correlation
approaches a near-constant value, and therefore, we use the ``derivative'' method proposed in our earlier work~\cite{Lin:2017ani} to obtain the quasi-PDF:
\begin{align}
\label{eq:derivative}
\Delta \tilde{q}_R(x,P_z) = \int_{-z_\text{max}}^{+z_\text{max}} \!\!\! dz \frac{ie^{i x P_z z}}{x} \partial_z\tilde{h}_R(z,P_z).
\end{align}
Again $\partial_z\tilde{h}_R(z, P_z)$ is consistent with zero for $|z|>15a$,
and we vary $z_\text{max}$ to estimate the error, which is small compared with other systematics.

\begin{figure}[htbp]
\includegraphics[width=.48\textwidth]{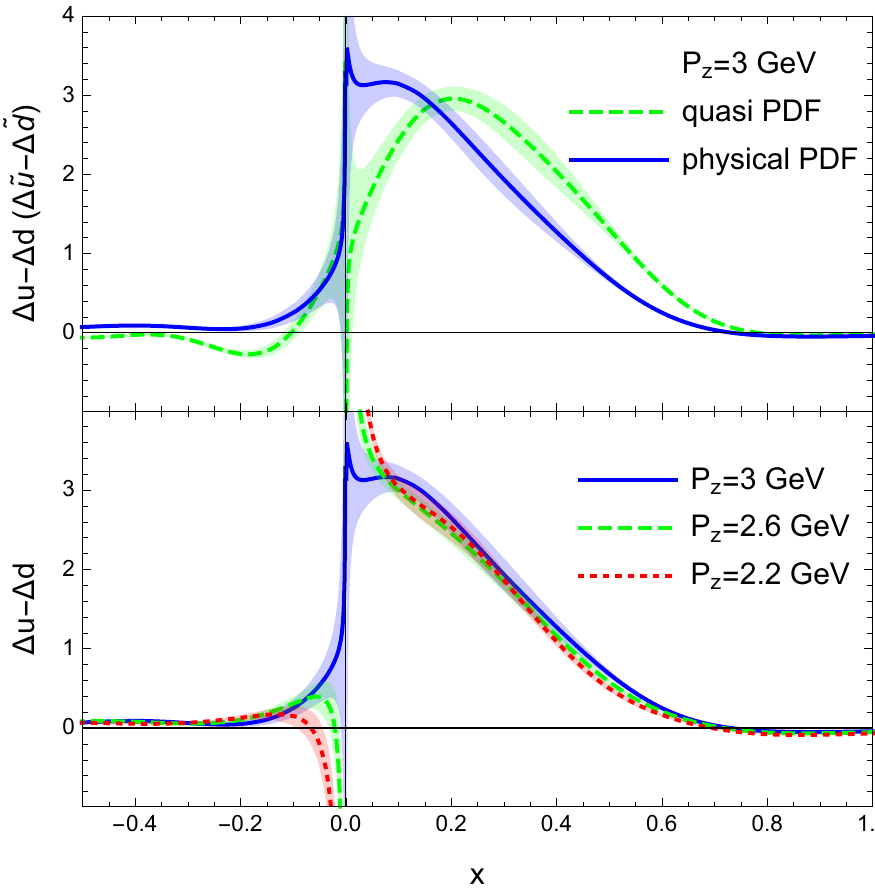}
\caption{The top panel is a quark helicity quasi-PDF in RI/MOM scheme at proton momentum 3.0~GeV
and resulting physical PDF in $\overline{\text{MS}}$ at $\mu=3$~GeV. The error bands are statistical.
The bottom panel shows the matched physical PDFs from various proton momenta.} \label{fig:matchingPDF}
\end{figure}

We show in the top panel of Fig.~\ref{fig:matchingPDF} a comparison between the renormalized quasi-PDF at $P_z=3.0$~GeV and
the isovector quark helicity distribution resulting from the matching formula in Eq.~\ref{eq:momfact} with proton-mass correction (see Ref. \cite{Liu:2018uuj} for details on the de-convolution).
The error bands are statistical only. The matching corrections suppress the distribution
at mid $x$ to large $x$, yielding a positive antiquark (negative-$x$ region) helicity for $x<-0.1$.
This is physically intuitive because matching is in some sense boosting the finite-momentum quasi-PDF
to an infinite-momentum one with proper renormalization, and boosting will in general move
large-$x$ partons to smaller $x$.
In the bottom panel, we show a comparison between the helicity distributions extracted from different
proton momenta. In the large-$x$ region, the
differences are small, indicating small higher-twist effects.
However, the central values at small and negative $x$ shift noticeably from 2.2 to 3.0~GeV, reflecting the change of the
limiting behavior of the lattice correlation $\tilde{h}(z,P_z,a)$ at large $zP_z$ shown in Fig.~\ref{fig:ZME}.

\begin{figure}[htbp]
\includegraphics[width=.48\textwidth]{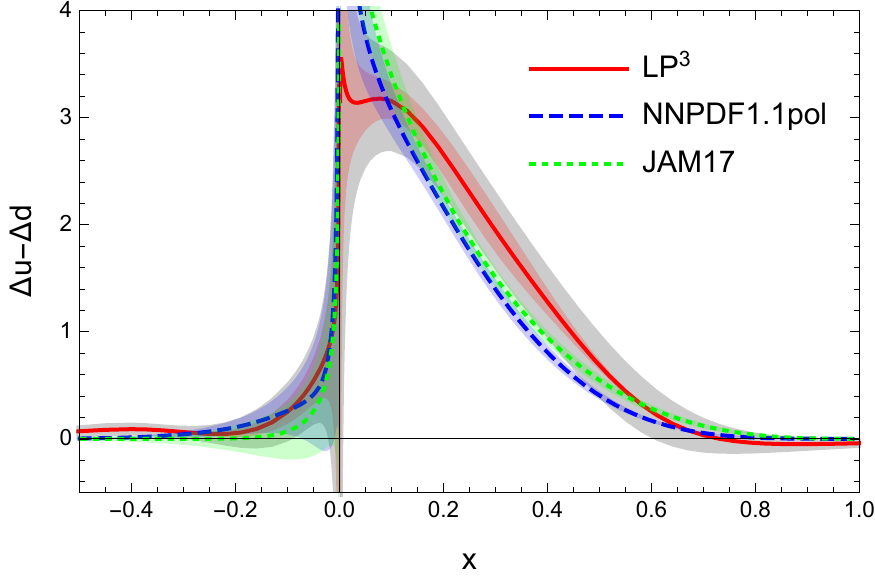}
\caption{The red line is the $\overline{\text{MS}}$-scheme isovector quark helicity PDF at
scale $\mu=3$~GeV, extracted from LaMET at the largest proton momentum (3~GeV),
compared with fits by NNPDFpol1.1~\cite{Nocera:2014gqa} and JAM~\cite{Ethier:2017zbq}. The red band contains
statistical error while the gray band also includes estimated systematics from finite lattice spacing, finite
volume, higher-twist corrections, as well as renormalization scale uncertainties.
} \label{fig:finalPDF}
\end{figure}

Our final isovector quark helicity distribution, obtained at the largest proton momentum of 3~GeV, is shown in Fig.~\ref{fig:finalPDF}.
The statistical error (with the excited-state contamination subtracted based on two-state fits) is shown as the red band.
The systematic uncertainty, shown combined in total with statistical one as the gray band in Fig.~\ref{fig:finalPDF},
is obtained partly by varying the scales in the NPR for $\mu_R \in \{2.3,3.7\}$~GeV and $p^R_z \in \{1.3,3\}$~GeV.
The error from one-loop matching inversion is estimated by the second-order correction.
The systematics associated with lattice spacing $a$ (discrete action,
mismatching in valence and sea fermions, and rotational symmetry violation, etc) and with
finite volume effects are estimated to be conservatively
about 8\% and 5\%, respectively, allowing a factor of 2-3 larger than the first-moment calculation itself
in Ref. \cite{Bhattacharya:2016zcn} to account for the unknown $x$-dependence and Lorentz-boost effect (see below).
The target-mass correction from Ref.~\cite{Chen:2016utp}
is found to be negligible for all three nucleon momenta,
again indicating small higher-twist contributions.
Also shown in the figure are the phenomenological
fits from  NNPDFpol1.1~\cite{Nocera:2014gqa} and JAM~\cite{Ethier:2017zbq}. The present
calculation is consistent with experiment within 1$\sigma$ in the large-$x$ region. For $x$
very close to 1, the calculation is in principle limited by the finite lattice spacing effect at large $P_z$,
where the proton needs be resolved with a finer longitudinal scale because of Lorentz contraction.
However, the consistency of data at small $zP_z$ in Fig.~\ref{fig:ZME} indicates that moderate $P_z$
may be sufficient for an accurate result. For $x<0.1$, the present calculation is limited by the accuracy
of large-$zP_z$ data. As in experiment, determining the small-$x$ PDFs
requires large-momentum hadrons.

The present calculation shows the potential impact of lattice simulations combined with the LaMET approach
in determining PDFs. The JLab 12-GeV program is well positioned to make
large-$x$ determinations of polarized and unpolarized parton distributions, which are
extremely valuable to interpret large $P_T$ events at the Large Hadron Collider. Lattice
calculations at 10\% level will already be very useful in deciding the large-$x$
behavior, cross-checking with the experimental data.

To summarize, we report a state-of-the-art isovector quark helicity distribution using lattice-QCD simulations at physical pion mass with proton momentum as
large as 3~GeV. With high statistics, we combined multi-state analysis and multiple source-sink separations to remove excited-state contamination from our analysis; its error is reflected in our statistical uncertainty. We renormalize the nucleon matrix element using the nonperturbative RI/MOM renormalization, and perform the LaMET one-loop matching to convert quasi-distribution to physical distribution in the $\overline{\text{MS}}$ scheme.
An estimate of the systematic uncertainty introduced by the choice of scales in the nonperturbative RI/MOM renormalization and one-loop matching inversion, as well
as finite lattice spacing and volume is included in the
final analysis. Our final result is consistent with the global analyses done by NNPDF and JAM within theoretical errors.
Future directions will be to investigate finer lattice-spacing ensembles and to reach even higher proton momenta, so that we can push toward
smaller $x$ in advance of upcoming experiments such as at the EIC.


\section*{Acknowledgments}
We thank the MILC Collaboration for sharing the lattices used to perform this study. The LQCD calculations were performed using the Chroma software
suite~\cite{Edwards:2004sx}.
This research used resources of the National Energy Research Scientific Computing Center, a DOE Office of Science User Facility supported by the Office of Science of the U.S.
Department of Energy under Contract No. DE-AC02-05CH11231
through ALCC and ERCAP;
facilities of the USQCD Collaboration, which are funded by the Office of Science of the U.S. Department of Energy,
and supported in part by Michigan State University through computational resources provided by the Institute for Cyber-Enabled Research.
HL, RL, and YY are supported by the US National Science Foundation under grant PHY 1653405 ``CAREER: Constraining Parton Distribution Functions for New-Physics Searches''.
JWC is partly supported by the Ministry of Science and Technology, Taiwan, under Grant No. 105-2112-M-002-017-MY3 and the Kenda Foundation. LJ is supported by the Department
of Energy, Laboratory Directed Research and Development (LDRD) funding of BNL, under contract DE-EC0012704. XJ is partially supported by the U.S. Department
of Energy Office of Science, Office of Nuclear Physics under Award Number DE-FG02-93ER-40762. XJ and YSL are partially supported by Science and Technology Commission of Shanghai
Municipality (Grant No.16DZ2260200) and National Natural Science Foundation of China (Grant No.11655002).  JZ is supported by the SFB/TRR-55 grant ``Hadron Physics from
Lattice QCD'', and a grant from National Science Foundation of China (No.~11405104). YZ is supported by the U.S. Department of Energy, Office of Science, Office of Nuclear
Physics, from DE-SC0011090 and within the framework of the TMD Topical Collaboration.

\bibliographystyle{apsrev}
\bibliography{ref}

\end{document}